\documentclass[
 aip,
 jmp,
 amsmath,amssymb,
 reprint,
]{revtex4-1}

\usepackage{graphicx}
\usepackage{dcolumn}
\usepackage{bm}

\begin{document}
\preprint{AIP/123-QED}

\title[]{Atomic resolution STM in a cryogen free dilution refrigerator at 15 mK}

\author{A.M.J.den Haan}
 \email{arthur.denhaan@gmail.com}
\author{G.H.C.J. Wijts}
\author{F. Galli}
 \affiliation{Department of Interface Physics, Leiden University, Niels Bohrweg 2, 2333CA Leiden, the Netherlands}
\author{O. Usenko}
 \affiliation{Leiden Cryogenics, Kenauweg 11, 2331BA Leiden, the Netherlands}
\author{G.J.C. van Baarle}
\author{D.J. van der Zalm}
 \affiliation{Leiden Spin Imaging (LSI), J.H. Oortweg 21, 2333CH Leiden, the Netherlands}
\author{T.H. Oosterkamp}
 \affiliation{Department of Interface Physics, Leiden University, Niels Bohrweg 2, 2333CA Leiden, the Netherlands}

\date{\today}

\begin{abstract}
Pulse tube refrigerators are becoming more common, because they are cost efficient and demand less handling than conventional (wet) refrigerators. However, a downside of a pulse tube system is the vibration level at the cold-head, which is in most designs several micrometers. We implemented vibration isolation techniques which significantly reduced vibration levels at the experiment. These optimizations were necessary for the vibration sensitive Magnetic Resonance Force Microscopy experiments (MRFM) at milli-kelvin temperatures for which the cryostat is intended. With these modifications we show atomic resolution STM on graphite. This is promising for scanning probe microscopy applications at very low temperatures.
\end{abstract}

\maketitle

\section{Introduction}
Pulse tube (PT) refrigerators have become the standard for many low temperature applications\cite{Uhlig2}. The main advantages of a PT-cooler are the significant reduction of labor intensity of precooling the dilution refrigerator or experiment as compared to cryogen cooled (dilution) refrigerators, where helium needs to be refilled regularly. These liquid helium transfers from storage dewar to experimental dewar often require that the running experiments are interrupted. This is in particular the case for very sensitive techniques, like scanning probes. In addition, considering the steep global increase of helium scarcity\cite{Nuttall2012}, running the pulse tube is much less costly and does not depend on the quality or quantity of the helium supply.

Even though PT refrigerator systems become more available, most of the low temperature vibration sensitive measurements like low temperature atomic force microscopy (AFM) and scanning tunneling microscopy (STM) are still performed in conventional (wet) (dilution) refrigerators\cite{Davis, Wiesendanger, tartaglini, Otte}. The reason is that the pulse tube relies on a varying pressure between 7 and 22 bars\cite{PT415, deWaele2011}, resulting on the one hand in square wave-like, low frequency, kilonewton forces acting on the top parts of the cryostat, and on the other hand in kilohertz range acoustical vibrations due to the gas flow through the pulse tube regenerator and the flexible hoses connected to the rotary valve and expansion vessels.

This paper describes vibration isolation techniques as well as STM measurements in a commercially available cryogen free dilution refrigerator \cite{LeidenC, pulsetube} (cryostat). The cryostat has a base temperature of less than 10 mK and a specified cooling power of 650 $\mu$W at 120 mK, which translates to 5 $\mu$W at 10 mK \cite{Pobell}. The vibration isolation is optimized for ultra-sensitive SQUID-based magnetic resonance force microscopy (MRFM) experiments\cite{Vinante}. To test the performance of the vibration isolation, we replaced the MRFM-setup with an STM-setup.
To our knowledge, we show for the first time atomic resolution STM in a pulse tube cooled (cryogen free) dilution refrigerator.

\section{\label{vibiso}Vibration Isolation of the cryostat}
\begin{figure}
\includegraphics[width=8.5 cm]{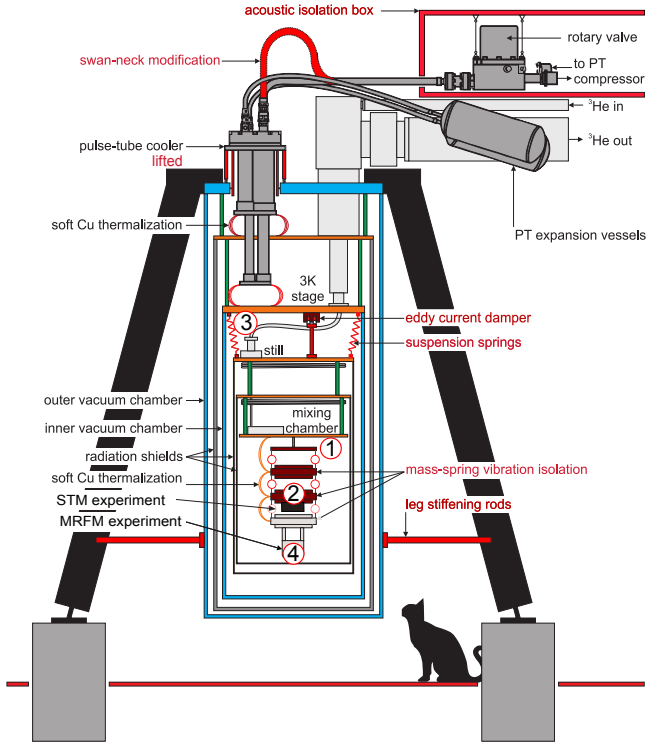}
\caption{\label{fig:cryostat} Schematic representation of the commercial cryogen-free dilution refrigerator\cite{LeidenC} with the implementation of vibration-reducing modifications, which are indicated by a red color. The various vibration measurements are performed at the positions indicated by the numbers 1 to 4 inside the red circles. Position 1: SQUID at the mixing chamber plate. Position 2: SQUID at the second mass. Position 3: Geophones at the 3 K plate. Position 4: MRFM vibration measurement inside MRFM-setup (aluminum box in the order of 10 x 10 x 10 cm)\cite{Wijts}.}
\end{figure}

In Fig.~\ref{fig:cryostat}, various modifications\cite{Wijts} to the factory default setup are shown that will be discussed in this paragraph.

The pressure variation in the pulse-tube (PT) is realized though a rotary valve that switches the PT-inlet between the 7 and 22 bar outputs of a compressor, with a frequency of 1.4 Hz. In order to reduce the horizontal forces acting on the cryostat, we have lengthened the hose between the PT and the rotary valve, implementing a flexible ``swan-neck" shape, and placed the rotary valve on a flexible platform inside an acoustic isolation box. The hoses between the rotary valve and the compressor are loosely suspended with ropes from the ceiling of an adjacent hallway. In this way, the expansion of the hose between PT and rotary valve results in an acceleration of the rotary valve and the hoses to the compressor, rather than that of the PT head\cite{Wijts}.

In the default configuration, the PT is rigidly connected to the plates at the room temperature-, 50 K-, and 3 K- stages. The periodic expansion of the PT due to the pressure variations is reported to be 25 $\mu$m \cite{Uhlig}. In order to reduce the forces acting between the top three cryostat stages, caused by this expansion, we have lifted the PT a few cm, so that it is resting on support rods and a rubber ring on the room temperature plate. The PT is thermalized to the 50 K and 3 K stages with soft copper braids\cite{Wijts}.

The three support legs of the cryostat are placed on a concrete block that is separated from the foundation of the surrounding building. We have stiffened the connection between the legs and the outer vacuum chamber (OVC) by making the leg connections to room temperature plate more bulky as well as by adding rods between the bottom of the OVC and the legs, thus creating triangular support structures.

The default connection between the 3 K plate and the still plate is provided by rigid poles. We have removed these poles and suspended the lower three stages of the cryostat from the 3 K plate with tension springs\cite{springs}. The total mass of the suspended part is 55 kg. We use 5 pairs of springs with a stiffness of 1.31 N/m, which leads to an estimated vertical resonant frequency of 3 Hz. In order to reduce the vibration amplitude at this frequency, which is uncomfortably close to the second harmonic of the PT excitation, we implemented an eddy current damper that is thermalized at the 3 K plate\cite{Wijts}.

\begin{figure}
\includegraphics[width=8.5 cm]{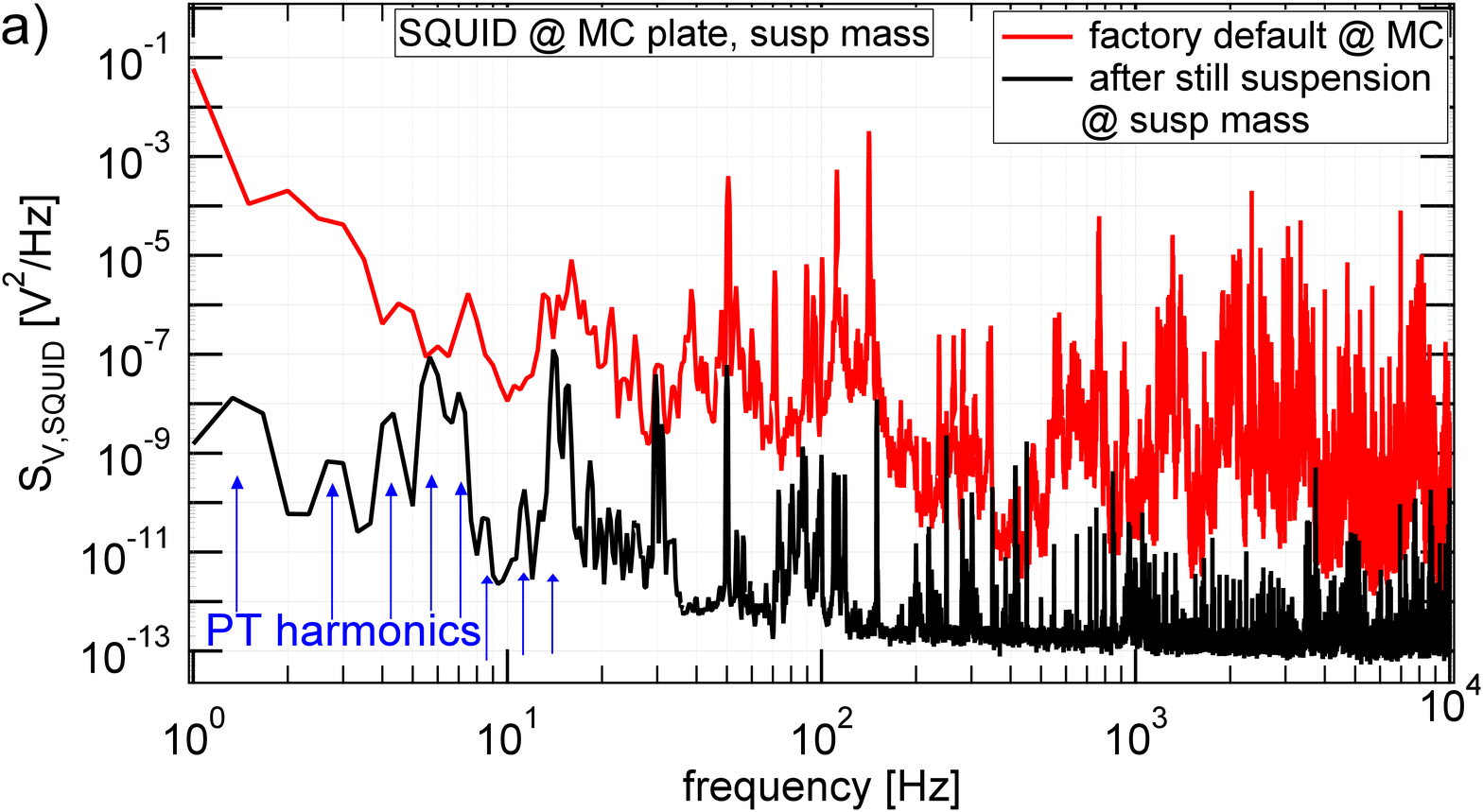}
\includegraphics[width=8.5 cm]{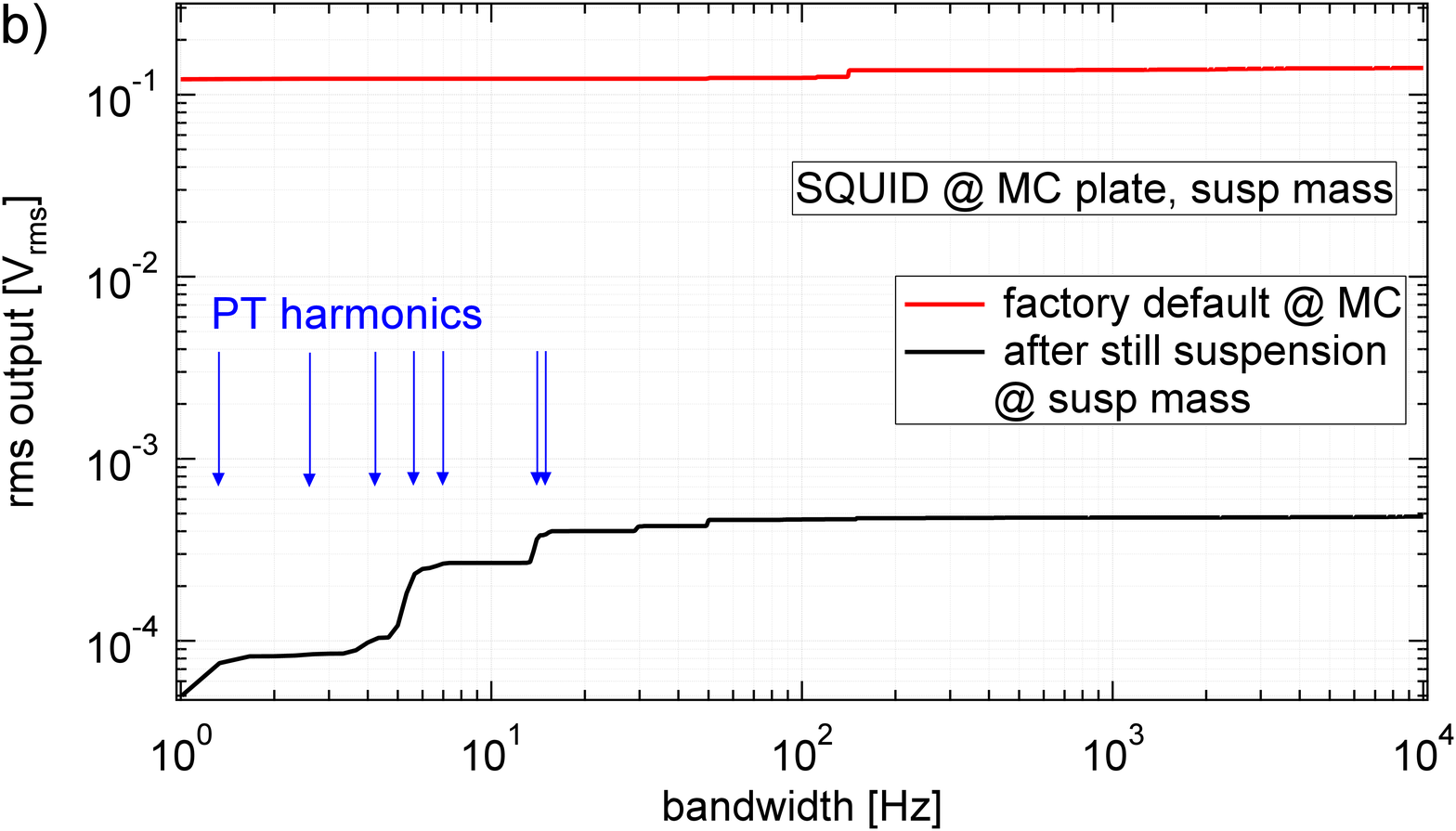}
\caption{\label{fig:suspension} Vibration reduction after suspension of the still plate from the 3 K plate and adding a mass-spring system below the mixing chamber. The SQUID  vibration measurements were performed at position 1 (factory default) and position 2 (after still suspension) in Fig.~\ref{fig:cryostat}. (a) PSD spectra of the SQUID noise\cite{Wijts} (b) The rms output: $\mathrm{SQUID output_{rms}}(\Delta f)=\sqrt{\int_{1}^{\Delta f} \! S_{V}(f) \, \mathrm{d}f}$. A significant vibration reduction over the whole spectrum is visible. The low frequency background noise is reduced to such a level that the harmonics of the PT modulation (n$\cdot$1.4 Hz) appear (peaks in the PSD spectrum and jumps in the integrated spectrum).}
\end{figure}

Furthermore, we added a mass-spring vibration isolation system below the mixing chamber plate, consisting of three 5 kg brass masses, suspended from tool steel ring springs, which was designed to provide a 100 dB vibration isolation in a frequency range between 1 kHz and 5 kHz \cite{Usenkophd}. The mass-spring vibration isolation is especially suitable for the MRFM cantilevers which have their natural resonance frequencies in this frequency range. The masses are thermalized to the mixing chamber with commercial soft Cu tape, clamped with brass bolts. The MRFM experiments were always mounted on the lowest or second-to-lowest mass of this mass-spring system. In the case of the STM measurement described in section \ref{sec:STMsetup}, the STM setup was mounted on the second mass, removing the third (lowest) mass, since vibration isolation in this frequency range (1 kHz to 5 kHz) is not necessary for STM. In order to measure the vibrations, we used a Superconducting QUantum Interference Device- (SQUID) based readout from our MRFM setup, which measures tiny magnetic flux changes \cite{Usenko2011}. The signals are resulting from the motion of the detection coil in the ambient magnetic field gradient that exists inside the cryostat\cite{Wijts}.

Fig.~\ref{fig:suspension}a shows the improvement in SQUID noise due to the suspension of the still plate and the implementation of the mass-spring vibration isolation. This was an early modification, before lifting the PT and adding the leg stiffening rods. The red curve in Fig.~\ref{fig:suspension}a shows the Power Spectral Density (PSD) of the SQUID voltage noise with the SQUID on the mixing chamber plate (position 1 in Fig.~\ref{fig:cryostat}) in the cryostat as delivered (but with the rotary valve suspension and the cryostat placement already as drawn in Fig.~\ref{fig:cryostat}). The black curve was measured with the SQUID on the second suspension mass (position 2 in Fig.~\ref{fig:cryostat}), after implementation of the still suspension (suspension springs and eddy current damper). Although we did not calibrate for a displacement measurement, it becomes clear from the spectra that the suspension systems lead to a tremendous reduction of the vibrations over the whole measurement bandwidth. Above 200 Hz, the noise floor is now determined by the intrinsic detection SQUID flux noise instead of experimental vibrations. Below 10 Hz, the harmonics of the PT modulation (n$\cdot$1.4 Hz) emerge that were first obscured by low frequency background noise. Taking the square root of the integrated spectra, we obtain the root mean square (rms) output in Volt, which is shown in Fig.~\ref{fig:suspension}b. The spectra are integrated from 1 Hz to the given bandwidth (horizontal axis). The data below 1 Hz is influenced by the DC offset of the SQUID output, which is therefore not integrated. We find a relative improvement of a factor 291 in a bandwidth from 1 Hz to 10 kHz by dividing the last point from the red curve (factory default) by the last point of the black curve (after suspension).

\begin{figure}
\includegraphics[width=8.5 cm]{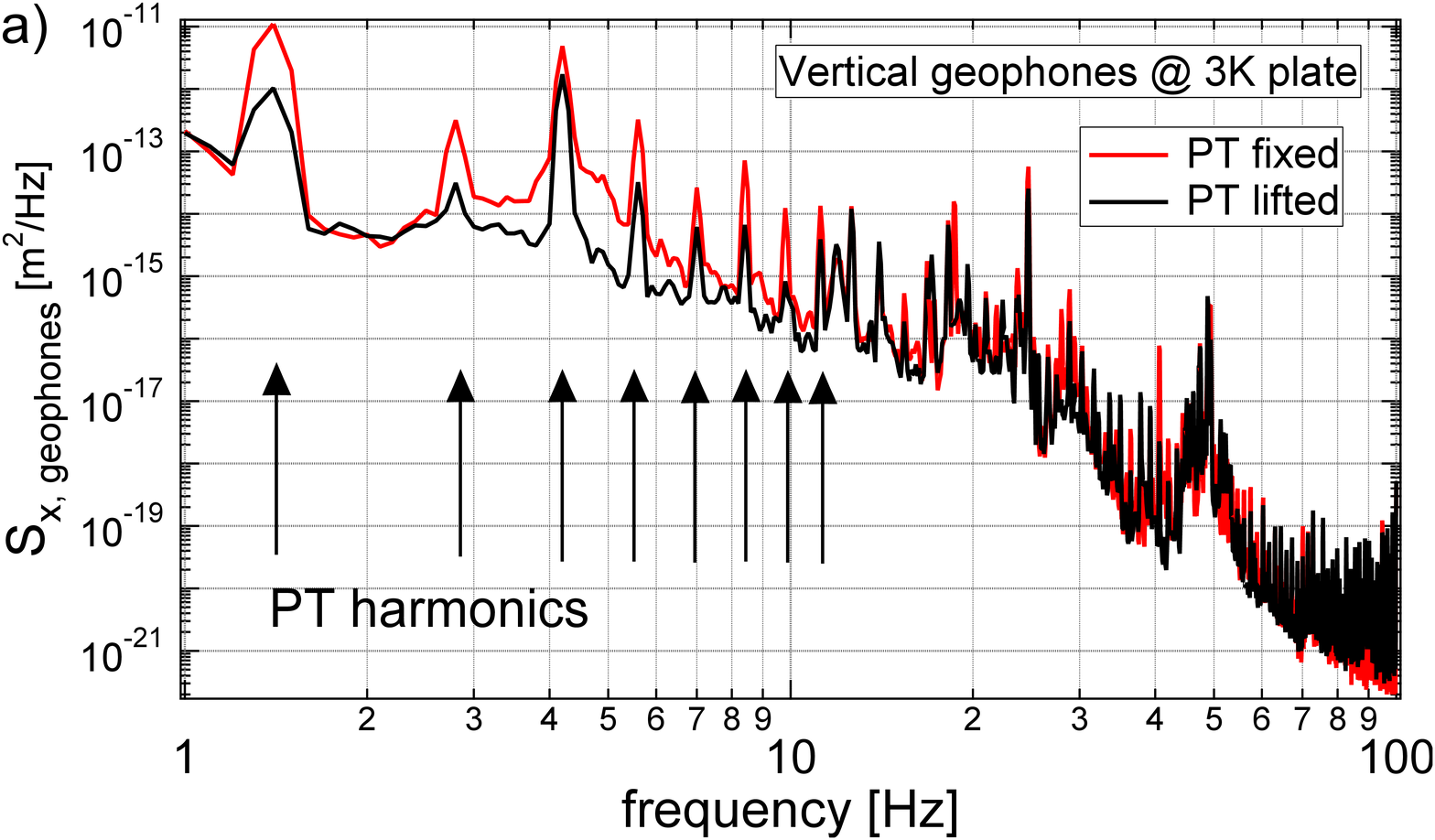}
\includegraphics[width=8.5 cm]{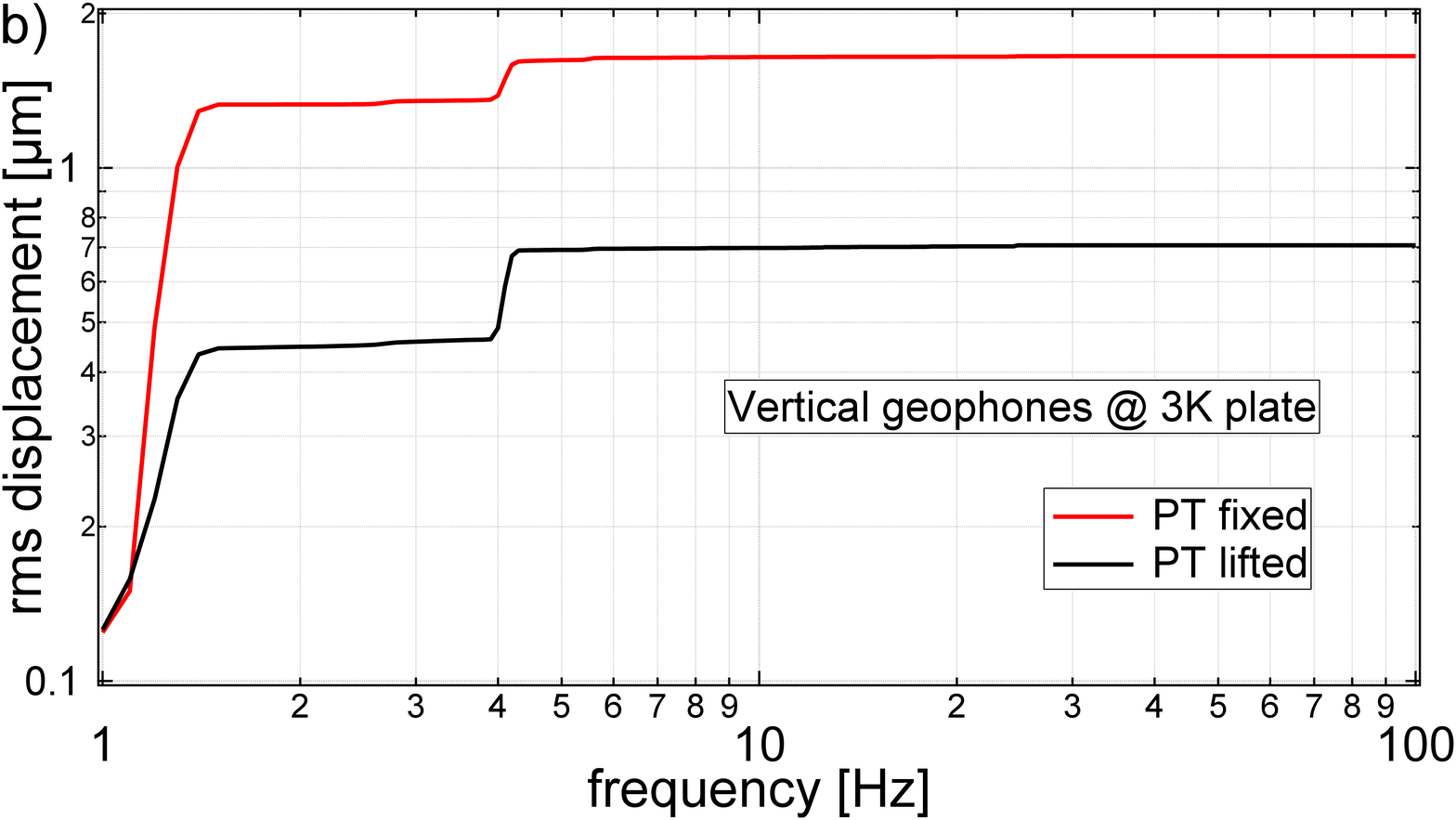}
\caption{\label{fig:PT} Vibration reduction after lifting the pulse-tube cooler\cite{Wijts}, measured at position 3 (3 K plate). (a) PSD spectra of geophone signals. (b) The rms displacement noise: $\mathrm{displacement_{rms}}(\Delta f)=\sqrt{\int_{1}^{\Delta f} \! S_{x}(f) \, \mathrm{d}f}$. A significant vibration reduction at the base frequency of the PT(1.4 Hz) is visible. The vibration level at the 3$^{rd}$ harmonic (4.2 Hz) of the PT is less influenced by the lifting.}
\end{figure}

The effect of lifting the PT cooler was quantified by measuring the vibrations at the 3 K plate (position 3 in Fig.~\ref{fig:cryostat}) with rotating-coil geophones\cite{geophones}. The geophones are calibrated for vertical displacements in a frequency range between 1 Hz and 100 Hz. In Fig.~\ref{fig:PT}a we show the PSD of the displacement noise before and after lifting the PT. The vibration peaks at multiples of the PT modulation frequency are clearly reduced up to the 8th harmonic. Taking the square root of the integral of these spectra, we obtain the displacement noise, shown in \ref{fig:PT}b. From this figure, we see that the total root mean square (rms) displacement noise before and after lifting the pulse tube is 1.65 $\mu$m and 0.71 $\mu$m respectively. Therefore, the full-bandwidth relative improvement is a factor 2.3\cite{Wijts}. Considering the ~300 times improvement at the mixing chamber plate and the 0.71 $\mu$m displacement noise at the 3 K plate, we estimate a displacement noise of roughly 3 nm at the second mass (position 1 in Fig.~\ref{fig:cryostat}) after lifting the pulse tube and still suspension.

\begin{figure}
\includegraphics[width=8.5 cm]{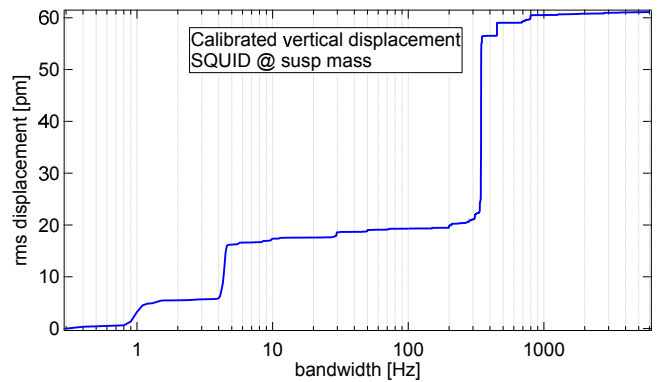}
\caption{\label{fig:allchanges} Net relative motion of tip versus sample in aluminum box at position 4 (inside the MRFM) after all modifications\cite{Wijts}.}
\end{figure}

The latest modification was to add rods between the OVC and the cryostat support legs in order to make the construction stiffer. In the original design, the connection between the support legs and the room temperature plate was quite floppy. The stiff cylindrical OVC connected to this plate acted like a vibration antenna that seemed to resonate with the 3rd harmonic of the PT modulation at 4.2 Hz. If we compare the PSDs of the displacement of a detection coil at the bottom suspension mass, with and without the extra rods, we observe that the peak at 4.2 Hz is reduced by a factor 2.5 (figure not shown)\cite{Wijts}. This increases however some vibrations at some other frequencies, most clearly at 1 Hz and at 342 Hz, but that still results in an improvement of a factor 1.2 when integrated over the whole measurement bandwidth\cite{Wijts}.

In Fig~\ref{fig:allchanges}, we have plotted the rms displacement noise of the detection coil with respect to the MRFM setup\cite{Vinante2} (position 4 in Fig.~\ref{fig:cryostat}) versus the integration bandwidth after the implementation of all modifications. The total rms displacement noise inside the MRFM setup is only 61 pm, which is more than sufficient to enable atomic-scale imaging. This value depends on the particular chosen setup, corresponding to different mechanical loops.

\section{\label{sec:STMsetup}STM setup and experimental preparation}
For these experiments we used an STM head built in-house similar to other setups used previously at low temperatures \cite{kelly, tartaglini}. The STM is based on the Pan walker design \cite{Davis, panpatent}. The coarse approach motor consists of a hexagonal cross-section prism made of titanium, covered by a hard coating (TiCN). The prism is held by three pairs of shear mode piezoelectric actuators. Two pairs are rigidly glued to the STM body, while the remaining pair is glued to a plate and pressed against the prism with a phosphor-bronze leaf spring. The normal force (and therefore the friction) between the actuators and the prism is regulated by adjusting the spring. The latter will also provide constant normal force if the size of the STM assembly changes because of thermal shrinkage at low temperatures.
The motor is driven by sending de-phased voltage signals to the pairs independently (walker mode) as described in ref. \onlinecite{Davis}.

The scan is performed by XY shear- and Z thickness mode piezoelectric elements \cite{piezo}. The feedback control and image acquisition are done by Leiden Probe Microscopy \cite{LPM} control electronics. Images were acquired in constant current- and constant height mode.

For the STM benchmark experiment, we used Highly Ordered Pyrolytic Graphite (HOPG), a standard STM sample for atomic resolution performance tests. The sample was glued on a 1 cm circular copper plate using low temperature compatible silver paint. The copper plate was mounted on an equivalent copper plate, using electrically isolating black Stycast, which served as a ground plate. Graphite was cleaved using scotch tape until a smooth surface was visible with the naked eye.
Furthermore, we used a platinum iridium tip, which was cut from a wire under a 45 degree angle. The sample and tip were thermalized to the 10 mK plate by using silver coaxial wiring. The total length of the wire from the tip to the current to voltage converter (gain 10$^{9}$) was 3 meter, with a capacitance of 240 pF. The STM was put at the 2nd mass below the 10 mK plate in a horizontal direction. In this direction, the STM is less sensitive to the Z (vertical) vibrations, which is the dominant vibration of the cryostat.

\begin{figure}
\includegraphics[width=8.5 cm]{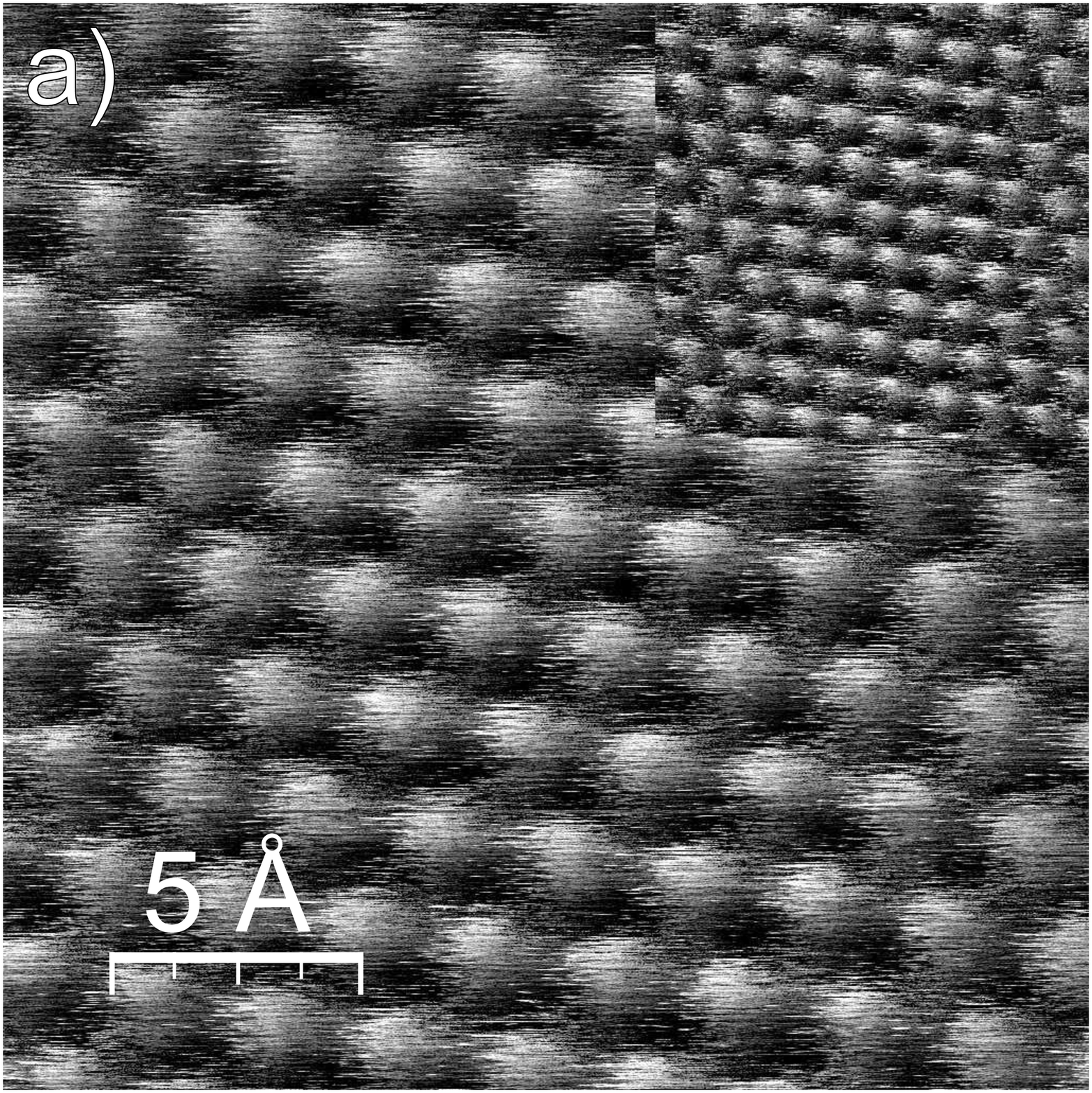}
\includegraphics[width=8.5 cm]{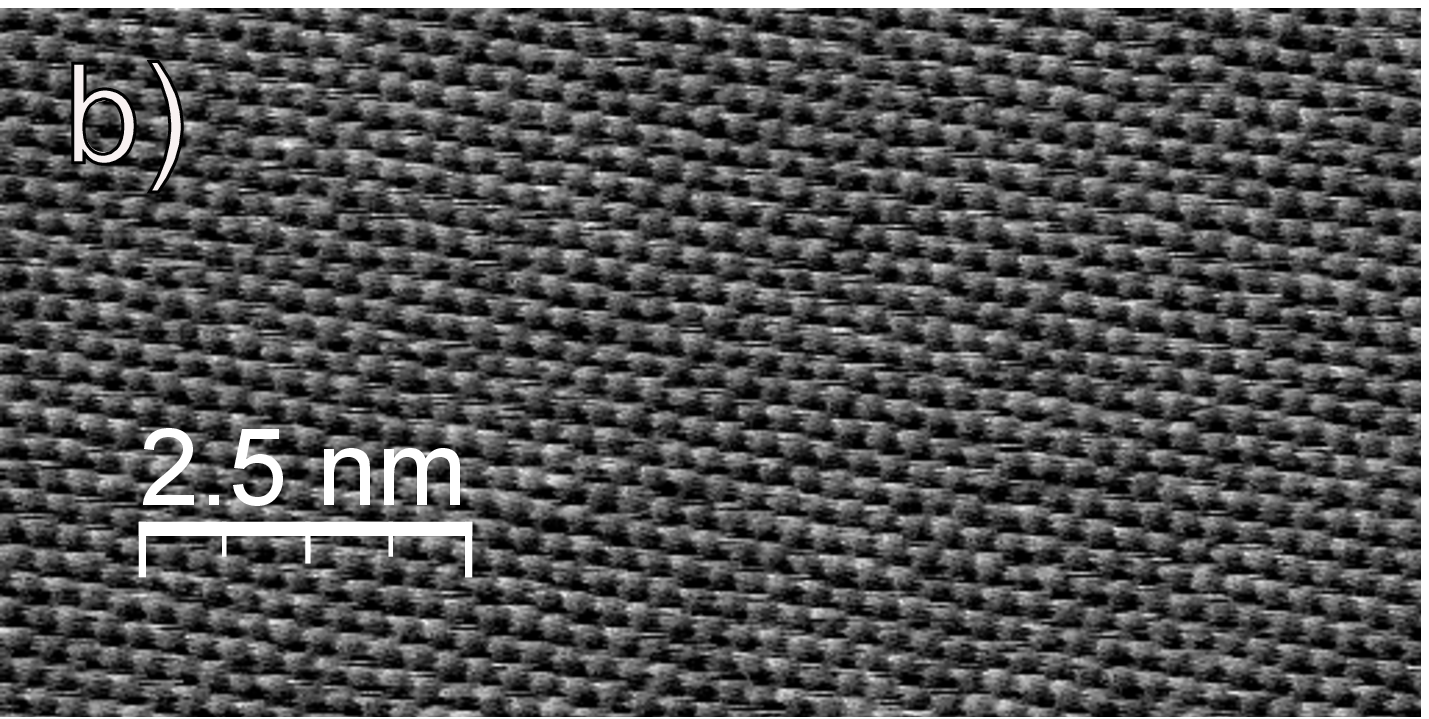}
\caption{\label{fig:STM} (a) constant current STM image (2.14 nm x 2.14 nm), showing atomic resolution on HOPG (left to right scan), the right to left scan image is shown in the inset. Frame time $T_{f}$ = 1049 seconds (1024x1024 pixels), tunneling current $I_{c}$ = 400 pA  and bias voltage $V_{b}$ = 0.5 V (b) constant current STM image (10.7 nm x 5.35 nm). $T_{f}$ = 262 seconds (512x256 pixels), $I_{c}$ =  400 pA and $V_{b}$ = 0.5 V.}
\end{figure}

\begin{figure}
\includegraphics[width=8.5 cm]{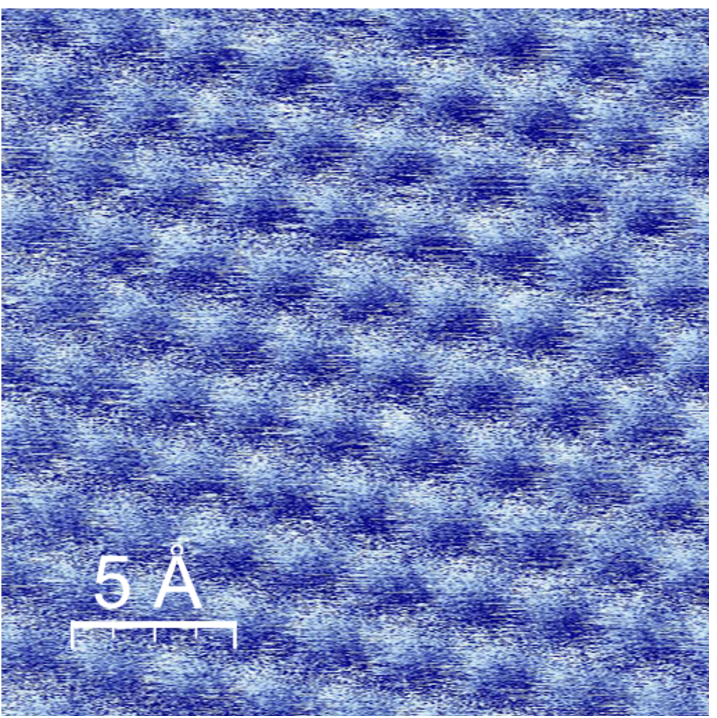}
\caption{\label{fig:STMtun} constant height image (2.14 nm x 2.14 nm). $T_{f}$ = 262 seconds (1024 x 1024 pixels), $I_{c}$ = 400 pA and bias voltage $V_{b}$ = 0.5 V.}
\includegraphics[width=8.5 cm]{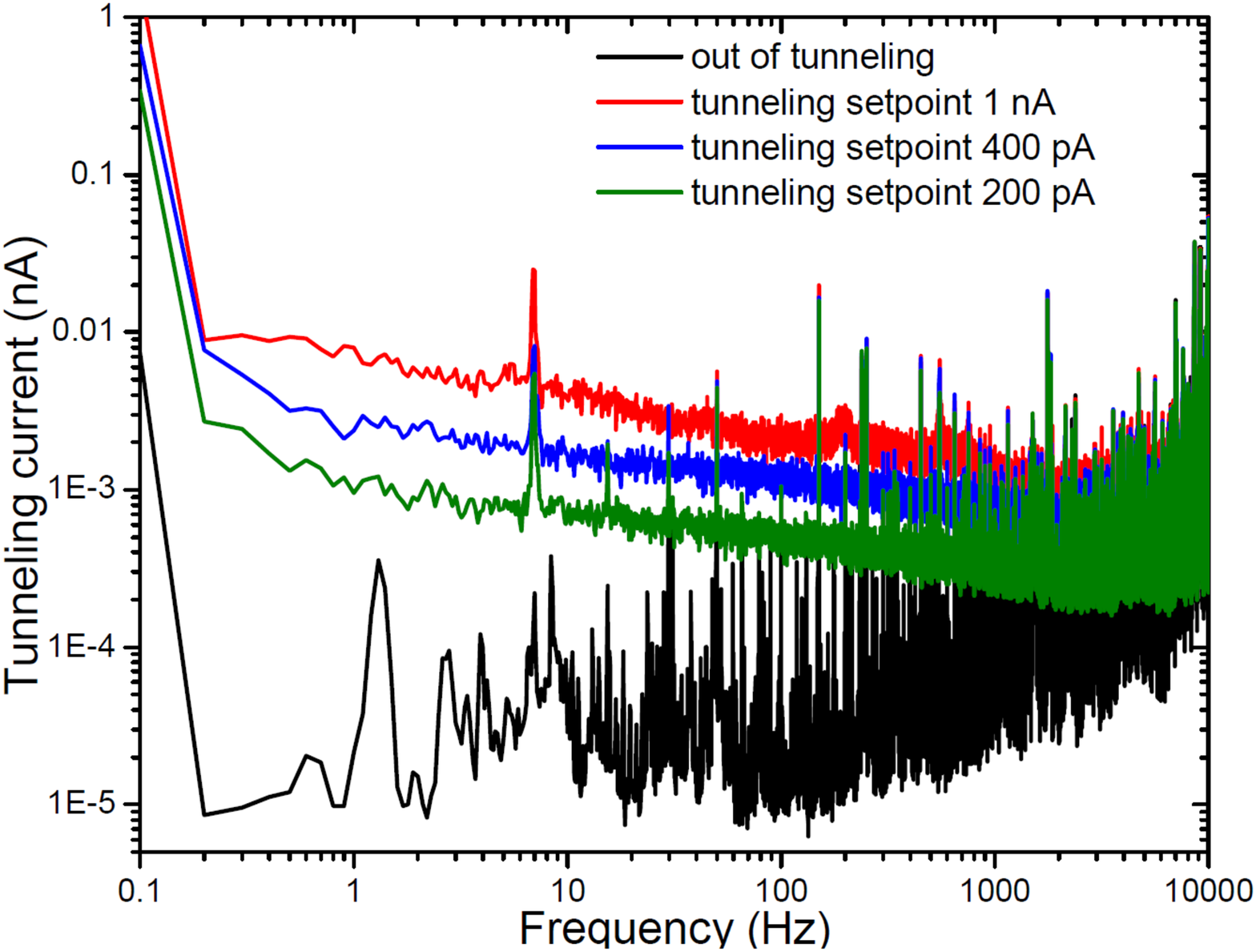}
\caption{\label{fig:STMspec} Tunnel-spectrum for several tunneling current setpoints. Most of the noise (peaks) visible in the out of tunneling spectrum are due to microphonics.}
\end{figure}

\section{Results}
In this section, we show several results, which serve as a benchmark for the vibration isolation of the pulse tube cooled dilution refrigerator described in section \ref{vibiso}. All data presented in this section is acquired with a running PT and dilution refrigerator at temperatures below 15 mK.
In Fig.~\ref{fig:STM}, two images in constant current mode are shown. The triangular atomic structure is clearly visible\cite{Hembacher}, showing that high resolution scanning probe measurements (SPM) are possible in cryogen free (dilution) refrigerators. The piezo-electric actuators (X, Y and Z) were otherwise not calibrated, but the scale of the atom separation from literature corresponds to the scale derived from the maximum displacement of the X,Y shear piezo’s at low temperatures. Furthermore, the image of the left to right scan is the same as the right to left scan (see the inset of Fig.~\ref{fig:STM}a).

We also measured at constant height, in which the current variations are measured while the voltage on the Z-piezo is kept constant to a certain extend. In the frozen state, the voltage drifted away from its initial value due to droop in the electronics, but this droop was slow enough to capture at least 2 images in 524 sec. The voltage drift resulted in a tip movement pointing away from the surface. In Fig.~\ref{fig:STMtun}, an atomic resolution image in constant height mode is shown. This image is a clear demonstration of stable tip surface situation, because present vibrations are not compensated by the feedback system in constant height mode.

In order to monitor the movement of the sample in comparison to the tip (thermal/piezo drift), we took several images in constant current mode of the same spot in 3 hours time, with a frame time of 1049 seconds. The frames of these measurements are compiled into a video, which is available in the supplementary information. One of the frames is shown in Fig.~\ref{fig:STM}a. From the displacement of the last frame compared to the first frame, we find a (X,Y)-drift of less than 1 {\AA} per 3 hours. This small drift and stable tip surface situation is promising for application in Scanning Tunneling Spectroscopy (STS) experiments.

In Fig. \ref{fig:STMspec}, the spectra of the tunneling current at several tunneling current setpoints are shown in constant height mode.

In the spectrum of the out of tunneling signal, several peaks are visible which in part corresponds to the pumps (30 Hz) and the pulse tube (1.4 Hz and higher order). These frequencies couple into the circuit due to microphonics. Using headphones to monitor the output of the current to voltage converter, the pulse tube, pumps and dilution refrigerator are very audible.
Note that the noise level increases when the current setpoint is raised in constant height mode, which is more dominant at lower frequencies. We attribute this effect to the voltage noise of a voltage source, connected to the Z-piezo, resulting in an excess piezo motion of an estimated 10$^{-12}$ $\mathrm{m/\sqrt{Hz}}$. In all of the spectra, a dominant peak at 6.9 Hz is clearly visible, which is the 5th harmonic of the pulse tube.

\section{Conclusion}
We significantly reduced the vibration level of a commercially available cryogen free pulse tube dilution refrigerator. The results show that future scanning probe microscopy experiments and other vibration sensitive experiments inside a cryogen free PT (dilution) refrigerators have become more accessible. Further vibration isolation will be implemented in future\cite{LSI}.

\begin{acknowledgments}
The authors would like to thank G. Koning from the fine mechanics department for his work in vibration improvement measures. They also would like to thank F. Tabak for providing the graphite samples and M. Rost and G. Verbiest for the use of their electronics. This work was supported in part by Fundamenteel Onderzoek der Materie (FOM), by the European Research Council (ERC) and by teh European project microKelvin.
\end{acknowledgments}

\providecommand{\noopsort}[1]{}\providecommand{\singleletter}[1]{#1}%

\end{document}